\documentclass[
aps,
 amsmath,
 amssymb,
reprint,
superscriptaddress,
floatfix,
]{revtex4-1}

\usepackage{graphicx}
\usepackage{dcolumn}
\usepackage{bm}

\usepackage{xcolor}
\usepackage{booktabs}
\usepackage{wasysym}
\usepackage[uncertainty-mode = separate]{siunitx}
\usepackage[utf8]{inputenc}
\usepackage[T1]{fontenc}
\usepackage{etoolbox}
\usepackage{hyperref}
\usepackage{stmaryrd}

\hypersetup{
    colorlinks = true,
    linkcolor = blue,
    citecolor = blue,
    urlcolor=blue,
}

\newcommand{\blue}[1]{\textcolor{blue}{#1}} 

\usepackage{tabularx}
\usepackage{appendix}
\usepackage{caption}
\makeatletter
\def\@email#1#2{%
 \endgroup
 \patchcmd{\titleblock@produce}
  {\frontmatter@RRAPformat}
  {\frontmatter@RRAPformat{\produce@RRAP{*#1\href{mailto:#2}{#2}}}\frontmatter@RRAPformat}
  {}{}
}%
\makeatother

\usepackage[font=small,labelfont=bf,format=plain]{caption}

\usepackage{setspace}
  \usepackage{parskip}
\usepackage[compact]{titlesec}         
\titlespacing{\section}{4pt}{4pt}{4pt} 
\AtBeginDocument{
  \setlength\abovedisplayskip{4pt}
  \setlength\belowdisplayskip{4pt}
  }
  
\usepackage[]{caption, subfig}
\usepackage{lineno}
\setcitestyle{square}
\usepackage{newunicodechar}
\newunicodechar{̈}{"{}}

\begin{document}

\newcommand{\ra}[1]{\renewcommand{\arraystretch}{#1}}
\newcommand{\deriv}{\mathrm{d}}

\title{Modeling the effect of MHD activity on runaway electron generation during SPARC disruptions}
\author{R. Datta}
\thanks{rdatta@mit.edu}
\affiliation{Plasma Science and Fusion Center, Massachusetts Institute of Technology, MA 02139, Cambridge, USA\looseness=-10000 
}%

\author{C. Clauser}
\affiliation{Plasma Science and Fusion Center, Massachusetts Institute of Technology, MA 02139, Cambridge, USA\looseness=-10000 
}%


\author{N. Ferraro}
\affiliation{Princeton Plasma Physics Laboratory, P.O. Box 451, Princeton, NJ 08543-0451, USA\looseness=-10000 
}%


\author{R. Sweeney}
\affiliation{Commonwealth Fusion Systems, Devens, MA 01434, USA\looseness=-10000 
}%

\author{R. A. Tinguely}
\affiliation{Plasma Science and Fusion Center, Massachusetts Institute of Technology, MA 02139, Cambridge, USA\looseness=-10000 
}%



\begin{abstract}

Magnetohydrodynamic (MHD) instabilities and runaway electrons (REs) interact in several ways, making it important to self-consistently model these interactions for accurate predictions of RE generation and the design of mitigation strategies, such as massive gas injection (MGI). Using \mbox{M3D-C1} – an extended MHD code with a RE fluid model – we investigate the effects of 3-D nonlinear MHD activity, material injection, and 2-D axisymmetric vertical displacement events (VDEs) on RE evolution during disruptions on SPARC – a high-field, high-current tokamak designed to achieve a fusion gain $Q > 1$. Several cases, comprising different combinations of neon (Ne) and deuterium ($\text{D}_2$) injection, are considered. Our results demonstrate key effects that arise from the self-consistent RE + MHD coupling, such as an initial increase in RE generation due to MHD instability growth, decreased saturation energies of the $m/n = 1/1$ mode driving sawteeth-like activity, RE losses in stochastic magnetic fields, and subsequent RE confinement and plateau formation due to re-healing of flux surfaces. Large RE plateaus (>5~MA) are obtained with Ne-only injection ($2\text{-}\SI{5e21}{}$ atoms), while combined $\text{D}_2$ + Ne injection ($\SI{2e21}{}$ Ne atoms; $\SI{1.8e22}{} \, \text{D}_2$ molecules) produces a lower RE current (<2~MA). With $\text{D}_2$ + Ne injection, a post thermal quench "cold" VDE terminates the RE beam, preventing a steady plateau. These simulations couple REs, 3-D MHD instabilities, MGI, and axisymmetric VDEs for the first time in SPARC disruption simulations and represent a crucial step in understanding RE generation and mitigation in high-current devices like SPARC.


\end{abstract}

\maketitle

\section{Introduction}
\label{sec:intro}

Runaway electron (RE) beams generated during disruptions in tokamaks pose a significant risk to plasma-facing components \citep{gill2002behaviour,granetz2014itpa,breizman2019physics,jepu2019beryllium,jepu2024overview,ratynskaia2025modelling,ataeiseresht2023runaway}. On high-current machines, small initial RE seeds can be amplified to Mega-Ampere-level RE beams, due to the exponential sensitivity of the RE avalanche gain to the pre-disruption plasma current \citep{breizman2019physics,martin2017formation,boozer2017runaway,vallhagen2020runaway}. RE mitigation, therefore, remains critical for future high-current tokamaks, such as SPARC \citep{creely2020overview,sweeney2020mhd}, ARC \citep{sweeney_arc_disruption2025}, and ITER \citep{martin2017formation,boozer2017runaway,vallhagen2020runaway}. 

SPARC, which is a compact ($a = \SI{0.57}{\meter},\, R_0 = \SI{1.85}{\meter}$) high-field ($B_0 = \SI{12.2}{\tesla}$), high-current ($I_p = \SI{8.7}{\mega\ampere}$) tokamak designed to achieve a fusion gain $Q > 1$, will experience both unmitigated and mitigated disruptions \citep{creely2020overview, sweeney2020mhd}.  SPARC is designed to be robust to the structural and heat loads generated during disruption events and will therefore serve as a test bed for how RE generation and mitigation strategies extrapolate to high-current machines \citep{sweeney2020mhd}. 

SPARC will be equipped with a 6 valve massive gas injection (MGI) system, which can deliver a combination of different gases for disruption mitigation \citep{sweeney2020mhd}. The baseline gas mix comprises deuterium ($\text{D}_2$) + neon (Ne) injection at a 9:1 ratio for simultaneous mitigation of thermal loads and RE generation \citep{sweeney2020mhd,kleiner2024extended,izzo2025disruption}.  DREAM \citep{hoppe2021dream} simulations of MGI-mitigated disruptions predict significant multi-MA RE plateau currents with Ne-only MGI, while additional $\text{D}_2$ injection is expected to limit the RE currents below $\SI{1}{\mega\ampere}$ in SPARC \citep{ekmark2025runaway}. 

SPARC will also deploy a runaway electron mitigation coil (REMC) --- a passively driven coil that seeds magnetohydrodynamic (MHD) instabilities during the current quench (CQ) \citep{sweeney2020mhd,tinguely2021modeling,tinguely2023minimum,battey2023design,izzo2022runaway}. The REMC relies on the enhanced transport of REs in stochastic magnetic fields to de-confine REs faster than they are generated \citep{smith2013passive}. REMC experiments on low-current devices such as J-TEXT \citep{li2023simulation,liu2025analysis,wei2025influence} and HBT-EP \citep{levesque2024design} have demonstrated promising results, while future implementation is planned on medium-current devices, such as TCV \citep{Sheikh2025Runaway}.

The effectiveness of the SPARC REMC was previously investigated by coupling the 3-D nonlinear MHD code NIMROD \citep{glasser1999nimrod} with the orbit-following code ASCOT \citep{hirvijoki2014ascot}, and the RE modeling framework DREAM \citep{hoppe2021dream,tinguely2021modeling,tinguely2023minimum}. These simulations demonstrated a plateau current of $I_f \approx \SI{5.5}{\mega \ampere}$ with Ne-only MGI ($\SI{4.8e21}{}$ atoms), while a complete suppression of the RE current was obtained with the REMC \citep{tinguely2021modeling}. Recent higher-fidelity NIMROD simulations, which include realistic MGI modeling and REMC vacuum fields, demonstrate a re-formation of closed flux surfaces after an initial period of stochasticity, which can be important for RE confinement \citep{izzo2025disruption}. 

 \begin{figure*}[t!]
\includegraphics[page=14,width=0.99\textwidth]{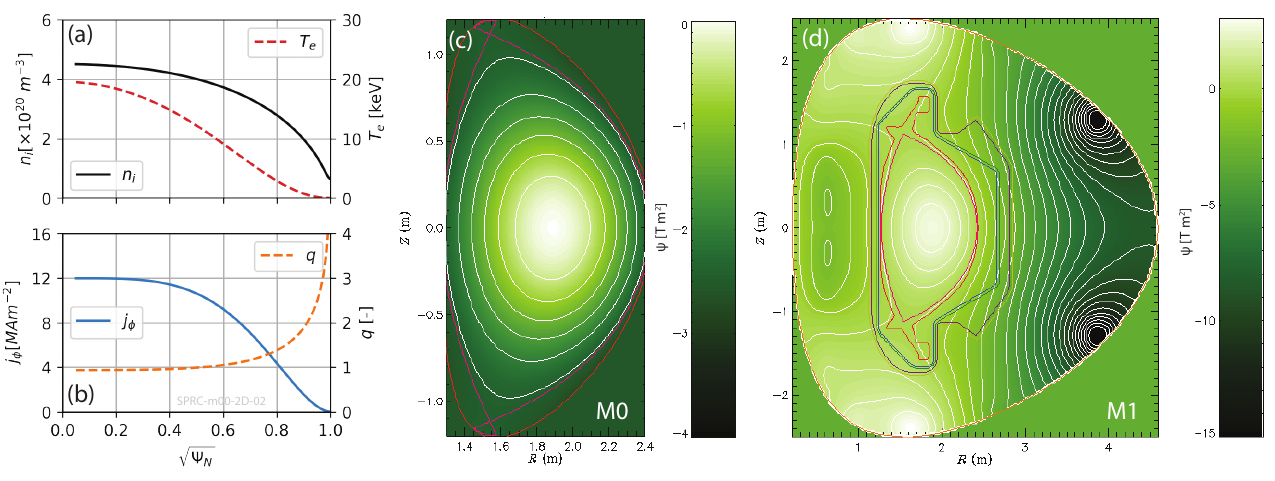}
\centering
\caption{ Flux-averaged profiles of the (a)~ion density $n_i$ (left) and electron temperature $T_e$ (right), and (b)~toroidal current density $j_\phi$ (left) and safety factor $q$ (right) at $t = \SI{0}{\milli\second}$. (c) M0 Simulation geometry and equilibrium distribution of the poloidal flux at $t = \SI{0}{\milli\second}$. White contours show lines of constant flux. The simulation geometry comprises the plasma volume (major radius $R_0 = \SI{1.85}{\meter}$, minor radius $a = \SI{0.57}{\meter}$) with a perfectly conducting (simplified) first wall. (d) M1 Double-walled vacuum vessel geometry comprising the first wall, inner vessel, outer vessel, and an outer vacuum region. A perfectly conducting boundary is imposed on boundary of the simulation domain. 
}
\label{fig:geometry}
\end{figure*}

 In addition to enhancing RE transport, MHD instabilities and REs can interact in several ways. For instance, peaked current profiles produced by RE generation can generate low on-axis safety factors, triggering MHD instabilities \citep{eriksson2004current,paz2019kink,matsuyama2017reduced,datta2025coupled,sainterme2024resistive,sainterme2025resistive}. Self-consistent simulations of RE and MHD physics have demonstrated increased linear growth rates of resistive kink and tearing-type modes \citep{helander2007resistive,zhao2020simulation,cai2015influence,liu2021self}, while a suppression of nonlinear sawteeth oscillations can also occur due to RE generation \citep{cai2015influence}. The final termination of a RE beam in a tokamak is typically attributed to external MHD modes or vertical displacement events (VDEs) \citep{yoshino2000runaway,paz2019kink,bandaru2021magnetohydrodynamic,bandaru2025axisymmetric,wang2024effect}. In some cases,  magnetic islands may also provide improved RE confinement \citep{papp2015energetic,yoshino2000runaway}. Some experiments \citep{chen2016enhancement} and reduced MHD + RE fluid simulations \citep{matsuyama2017reduced} have also shown an enhancement of RE currents due to MHD instabilities.

 The disruption modeling results for SPARC described earlier \citep{tinguely2021modeling,kleiner2024extended,ekmark2025runaway,izzo2025disruption} lacked a self-consistent coupling of the RE and MHD physics. Recently, self-consistent 2-D axisymmetric MHD + RE fluid calculations were performed with the extended MHD code {M3D-C1} \citep{jardin2012multiple} for SPARC disruptions \citep{datta2025coupled}. These simulations investigated the influence of different primary seeds on RE generation, and the final RE plateau and current profiles obtained in M3D-C1 were shown to be in good agreement with simpler 0-D/1-D models \citep{datta2025coupled}. These simulations, however, excluded 3-D MHD instabilities, impurity injection, and vertical motion of the plasma \citep{datta2025coupled}.
 
 In this paper, we complement previous modeling efforts with 3-D nonlinear MHD + RE simulations of SPARC disruptions, performed using {M3D-C1}, which couples a RE fluid model to the resistive MHD equations \citep{liu2021self,zhao2020simulation,datta2025coupled}. We model RE generation during the CQ for different quantities of neon (Ne) and deuterium ($\text{D}_2$) injection. Fusion-relevant primary RE sources, such as Compton scattering and tritium beta decay \citep{datta2025coupled}, are included in the present simulations. We also extend {M3D-C1}'s capabilities for modeling RE generation with impurity injection by implementing a partially screened neural network model for Dreicer generation \citep{hesslow2019evaluation}, not used in previous {M3D-C1} simulations. Note that the effect of the REMC is not included in the present work, and will be the subject of a future investigation. Our results demonstrate several important consequences of RE + MHD coupling, such as an enhancement of RE currents due to MHD activity, an abatement of sawtooth-like activity by RE generation, and RE losses due to magnetic stochasticity, before eventual re-healing of flux surfaces and plateau formation. We further simulate a post-thermal quench "cold" vertical displacement event (VDE) in 2-D with simultaneous RE generation in a disruption with combined $\text{D}_2$ + Ne injection. Here, the RE current grows to about $\SI{2}{\mega\ampere}$, before the vertical displacement terminates the RE beam, precluding plateau formation. Furthermore, a comparison of the VDE with and without REs demonstrates a slowdown of the vertical motion and later plasma termination with REs. The simulations presented in this paper self-consistently combine RE generation, MHD activity, and material injection for the first time in SPARC disruption simulations, guiding our understanding of RE generation and mitigation on compact, high-current tokamaks, such as SPARC.

\section{Simulation Setup}
\label{sec:methods}


\textit{Initial Conditions and Geometry –} The simulations in this paper are performed using the extended MHD initial value code {M3D-C1} \citep{jardin2012multiple}. In each simulation, the initial equilibrium comprises the primary reference discharge (PRD) in SPARC, which describes a double-null diverted, H-mode plasma with plasma current $I_p = \SI{8.7}{\mega\ampere}$, toroidal magnetic field $B_0 = \SI{12.2}{\tesla}$, and volume-averaged density and temperatures $\langle n_i \rangle \approx 3\times 10^{20} \SI{}{\per \meter \cubed}$ and $\langle T_e \rangle \approx \SI{7}{\kilo \electronvolt}$ \citep{creely2020overview,rodriguez2020predictions,sweeney2020mhd,tinguely2021modeling}. Although the SPARC PRD comprises a deuterium-tritium plasma, we model the primary ion species as purely deuterium in these simulations. The initial distributions of the flux-averaged ion density $n_i$, electron temperature $T_e$, toroidal current density $j_\phi$, and safety factor $q$ as a function of $\sqrt{\psi_N}$ are shown in \autoref{fig:geometry}\blue{(a-b)} \citep{rodriguez2020predictions}. Here, $\psi_N = (\psi - \psi_0)/(\psi_a-\psi_0)$ is the normalized poloidal flux, calculated from the flux at the magnetic axis $\psi_0$ and the plasma boundary $\psi_a$. The on-axis safety factor is below unity ($q_0 \approx 0.93$), and the $q = 1$ surface appears at roughly half the minor radius $\sqrt{\psi_N} \approx 0.5$.


\autoref{fig:geometry}\blue{c} shows the simulation geometry and the equilibrium distribution of the poloidal flux $\psi_P$ at $t = \SI{0}{\milli\second}$, which marks the pre-disruption flattop phase. The simulation geometry (M0) comprises the plasma volume (major radius $R_0 = \SI{1.85}{\meter}$, minor radius $a = \SI{0.57}{\meter}$) with a perfectly conducting (simplified) first wall. To determine the effects of wall resistivity and geometry, a limited number of simulations were performed with a more realistic (but computationally more expensive) SPARC geometry shown in \autoref{fig:geometry}\blue{d}. This double-walled vacuum vessel (VV) geometry (M1) comprises the first wall, inner vessel, outer vessel, and an outer vacuum region, with realistic resistivity values specified in each region ($\eta \approx \SI{8e-7}{\ohm\meter}$ in the inner/outer vessel, non-uniform parallel Spitzer resistivity in the plasma). The region between the first wall and inner vessel is not a toroidally conducting structure in SPARC and is modeled as a region of high resistivity ($\eta \approx 10^{-3}\SI{}{\ohm\meter}$). The boundary of outer vacuum region marks the boundary of simulation domain and is treated as a perfect conductor. Ports in the inner and outer vessel walls are also included in this geometry, which are modeled as regions of high resistivity in the toroidal direction ($\eta_\varphi \approx 10^{-3}\SI{}{\ohm\meter}$), and low resistivity in the poloidal direction ($\eta_{RZ} \approx \SI{8e-7}{\ohm\meter}$), similar to that in \blue{Ref. \citenum{kleiner2024extended}}. With the simplified conducting first wall geometry (M0), only the poloidal magnetic energy inside the first wall is available for RE generation, while with the M1 geometry, external magnetic energy may diffuse into the plasma on resistive time scales \citep{martin2015avalanche,datta2025coupled}. Here, roughly $\SI{50}{\mega\joule}$ of poloidal magnetic energy is available within the first wall, about $\SI{70}{\mega\joule}$ within the inner vessel, and $\sim \SI{150}{\mega\joule}$ in the entire simulation domain (excluding the contribution of the poloidal field coils). {The wall time is about $\tau_w \approx \SI{46}{\milli\second}$ \citep{kleiner2024extended}}. The phenomena modeled here exhibit time scales much faster than $\tau_w$; thus, we do not expect resistive effects to significantly affect the results. 


\begin{table}\centering
\ra{1.3}
\caption{
Quantity of injected material in \textsc{M3D-C1} simulations. 
$S \equiv [(\eta/\mu_0)\tau_A/a^2]^{-1}$ is the Lundquist number in the post-TQ plasma (at $t \approx \SI{0.1}{\milli\second}$), calculated using the flux-averaged resistivity $\eta$ at the $q = 1$ surface, and $\tau_A = R_0 / v_A = R_0 / (B_0/\sqrt{\mu_0 \rho})$ is the characteristic Alfvén time calculated using the toroidal magnetic field $B_0 = \SI{12.2}{\tesla}$ and the volume-averaged mass density $\rho$ (including deuterium and neon). The final column lists the on-axis parallel electric field (normalized by the Connor–Hastie critical field $E_C$) at $t \approx \SI{0.1}{\milli\second}$.%
}
\begin{tabular}{ccccc}
\hline
Case & Ne [atoms] & D\textsubscript{2} [molecules] & $S$ & $E_{\parallel,0}^*$ \\
\hline
(A) & 0 & 0 & \SI{7e5}{} & \SI{200}{} \\
(B) & \SI{4.8e21}{} & 0 & \SI{5e4}{} & \SI{60}{} \\
(C) & \SI{2e21}{} & 0 & \SI{1e5}{} & \SI{50}{} \\
(D) & \SI{2e21}{} & \SI{1.8e22}{} & \SI{2e5}{} & \SI{7}{} \\
\hline\hline
\end{tabular}
\label{tab:table}
\end{table}





{M3D-C1} solves the MHD equations in the region within the first wall using reduced quintic $C^1$ finite elements in the poloidal plane and cubic finite elements in the toroidal direction \citep{jardin2012multiple}. The simulation domain is discretized with an unstructured mesh comprising triangular elements in the poloidal plane and prism-shaped elements in the toroidal direction \citep{jardin2012multiple}. The 3-D simulations described here have 4 toroidal planes (enabling us to resolve $n<4$ toroidal modes). To probe the effect of higher $n$ modes, some simulations are performed with 8 toroidal planes, as described later in \blue{\S}\ref{sec:results}. 



\begin{figure}[b!]
\includegraphics[page=28,width=0.48\textwidth]{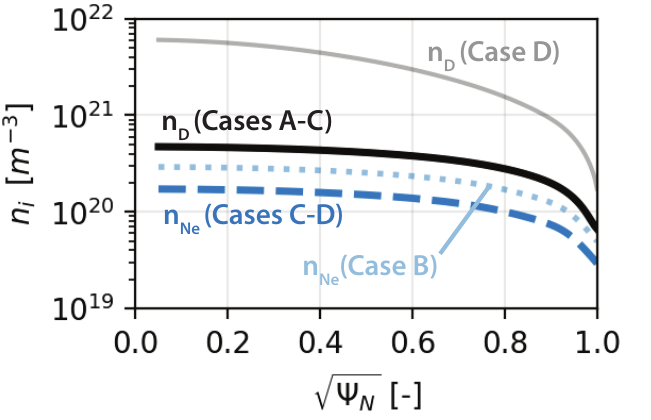}
\centering
\caption{Initial flux-averaged distributions of the main ion (D) and Ne densities. The solid black line shows $n_D$ for cases A-C and the gray solid line shows $n_D$ for case D (combined Ne + D$_2$ MGI). The dashed blue line shows $n_{Ne}$ for cases C-D, and the dotted blue line shows that for case B. There is no injected Ne in case A.
}
\label{fig:initial_ni}
\end{figure}

\textit{Injected Material –} In this paper, we model 4 different combinations of injected Ne and $\text{D}_2$; the injected quantities are summarized in \autoref{tab:table}. In case A, we assume a pure deuterium plasma, and do not directly model any impurities or additional injection. Cases B and C represent Ne-only injection. Here, we assume an initial distribution of neutral Ne at $t = \SI{0}{\milli\second}$ that scales linearly with the ion density profile $n_{Ne} \propto n_i$. This provides about \SI{4.8e21}{} Ne atoms in case B, similar to the level of Ne reported in \citeauthor{tinguely2021modeling}, while case C assumes about \SI{2e21}{} Ne atoms, closer to the assimilated level presently anticipated for SPARC. In the simulation with combined $\text{D}_2$ + Ne injection (case D), roughly $\SI{1.8e22}{}$ deuterium molecules are injected at $t = \SI{0}{\milli\second}$, providing a 9:1 ratio of $\text{D}_2$:Ne. This is the nominal baseline MGI ratio presently planned for SPARC, although other combinations are also possible. The SPARC MGI system is designed to deliver roughly $\sim 10^{22}$ Ne atoms and $\sim 10^{23}$ $\text{D}_2$ molecules to the plasma within a 2~ms gas pulse; thus, case D represents roughly $10\%$ assimilation of the delivered material \citep{ekmark2025runaway,izzo2025disruption}. For simplicity, both Ne and $\text{D}_2$ are assumed to be pre-distributed in the plasma. The initial distributions of Ne and D densities in the different cases are shown in \autoref{fig:initial_ni}. 


Once injected, Ne and $\text{D}_2$ are assumed to exhibit the same temperature $T_i$ as the primary ion species, become ionized, and are advected by the plasma motion. They further provide dilution cooling of the main ion and electron temperatures \citep{ferraro20183d}. The Ne ionization level and emitted radiation (which comprises line, bremsstrahlung, and recombination radiation) are calculated using the KPRAD module in {M3D-C1} \citep{ferraro20183d,kleiner2024extended}. The simplified initial distributions of Ne and $\text{D}_2$ are not fully representative of MGI in SPARC, where a 6-injector system will be employed \citep{kleiner2024extended,izzo2025disruption}. However, the present approach provides a reasonable starting point for determining how partially ionized impurities and low-Z injection affect the MHD and RE evolution, and more realistic injection at the location of MGI valves will be the subject of future research.

\begin{figure*}[t!]
\includegraphics[page=5,width=0.99\textwidth]{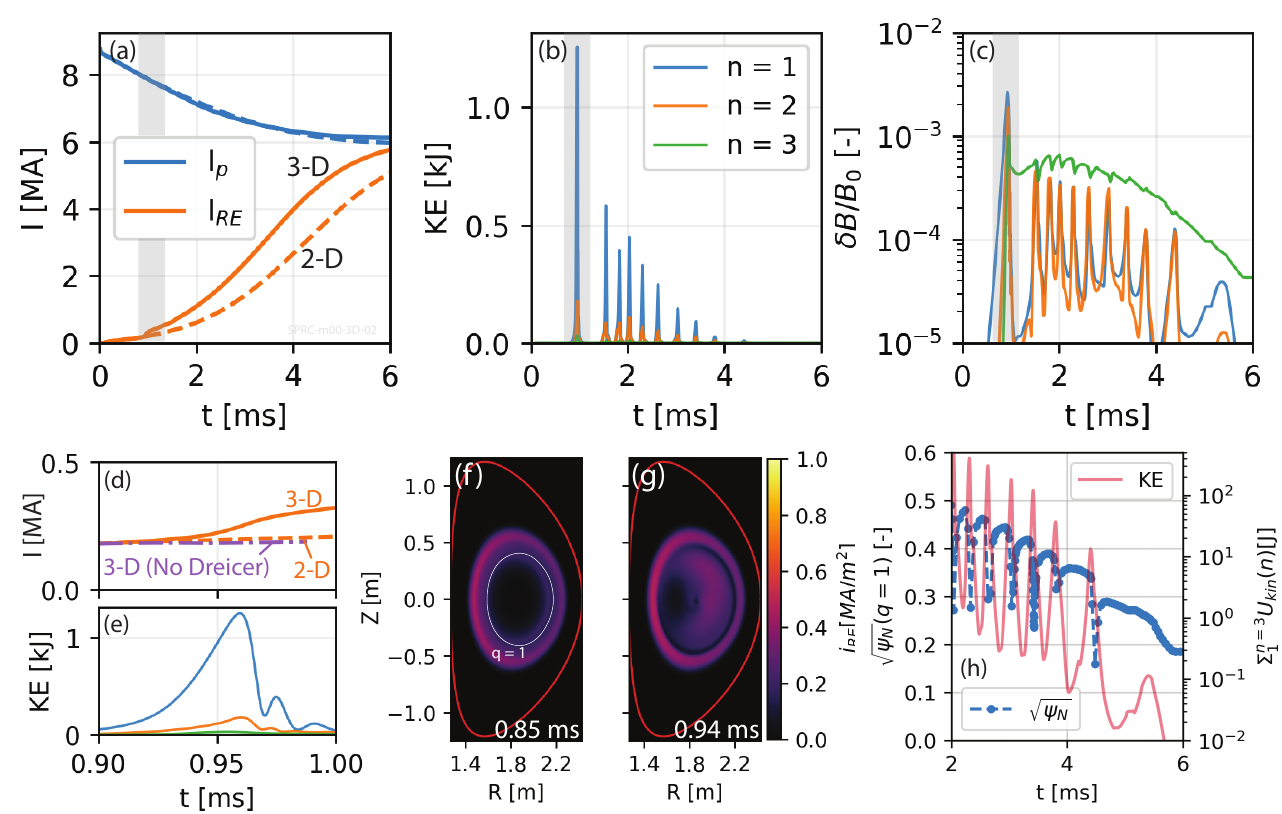}
\centering
\caption{ (a)~Evolution of plasma (blue, solid) and RE (orange, solid) currents in the 3-D nonlinear simulation of an idealized disruption without impurities (case A). The dashed lines represent the result from a 2-D axisymmetric simulation. (b)~Kinetic energy as a function of toroidal mode number $n$. (c)~Magnetic field line perturbation as a function of toroidal mode number $n$. (d-e)~Magnified version of \autoref{fig:unmitigated_3d}\blue{(a-b)} showing the runaway electron current evolution and the mode kinetic energy during the first crash. The purple line shows the result for the 3-D simulation where the Dreicer term is turned off around $t = \SI{0.9}{\milli\second}$. (f-g)~RE current density at $t = \SI{0.85}{\milli \second}$ and $t = \SI{0.94}{\milli \second}$ at the $\phi = 0$ plane. (h) Temporal evolution of the square root of the normalized poloidal flux $\sqrt{\psi_N}$ at the $q=1$ surface, as a function of time. The red line shows the total kinetic energy of the $1\leq n \leq 3$ modes (right axis). We plot values between $\SI{2}{\milli\second}< t < \SI{6}{\milli\second}$, during which the {M3D-C1} results were output at a higher frequency.
}
\label{fig:unmitigated_3d}
\end{figure*}


\textit{Runaway Electron Fluid Model –} We use a fluid runaway electron model, the {M3D-C1} implementation of which is described in detail in \blue{Ref.}~\citenum{zhao2024simulation}. Primary RE sources include Driecer generation \citep{dreicer1959electron}, Compton scattering \citep{martin2017formation}, and tritium beta decay \citep{martin2017formation}, while Rosenbuth-Putvinski avalanching \citep{rosenbluth1997theory} describes the secondary RE source. The implementation of these source terms is described in \blue{Ref.}~\citenum{datta2025coupled}. The tritium beta decay rate is calculated assuming that half the pre-MGI ion density consists of tritium. The Compton source is calculated from OpenMC calculations of the prompt gamma-flux $\Gamma(E_\gamma)$ inside the SPARC vacuum vessel due to the activated walls \citep{Wang2025ANOpenMC}. After $T_e$ has dropped below \SI{2}{\kilo\electronvolt} and the fusion rate becomes negligible, we artificially reduce the Compton rate by a factor of $10^{-3}$ to model the delayed non-prompt gamma-flux, similar to that in previous simulations \cite{ekmark_2025,ekmark2025runaway,datta2025coupled}. 

The presence of partially ionized Ne in the plasma is expected to reduce the Dreicer rate, due to partial screening of Coulomb interactions by bound electrons \citep{hesslow2019evaluation}. We therefore use a partially screened Dreicer model in these simulations, which we implement as a pre-trained neural network into {M3D-C1} \cite{hesslow2019evaluation}. An exception is case A, where we model a pure deuterium plasma, and thus, use canonical Dreicer generation \citep{dreicer1959electron}. Impurities and radiation can also raise the critical energy required for runaway formation and modify the avalanching rate \citep{hesslow2019influence,bandaru2024runaway}. However, these effects are not currently implemented in {M3D-C1} and will be the subject of a future publication. Hot-tail contribution is also not considered in the present work, and the initial RE current is set to zero. While hot-tail generation has been considered in previous RE modeling for SPARC using non-MHD codes such as DREAM \citep{tinguely2021modeling,ekmark2025runaway} and CODE \citep{sweeney2020mhd}, this has been done mainly using a kinetic approach where the two-dimensional RE momentum-space distribution function is evolved. In our {M3D-C1} simulations, a direct implementation of the hot-tail seed is not currently implemented and will be pursued in the future. 

\textit{Thermal quench} – Starting from the PRD initial conditions described previously (\autoref{fig:geometry}), we trigger an artificial thermal quench (TQ) by setting the perpendicular thermal diffusivity to a large value $\chi_\perp \sim 10^5\SI{}{\meter \squared \per \second}$ for $t < \SI{0.1}{\milli\second}$, similar to TQ initiation in various disruption studies \citep{clauser2019vertical,clauser2021iter,tinguely2021modeling,liu2021self}. We model the TQ using a 2-D axisymmetric {M3D-C1} simulation.  This is done to circumvent the numerical challenges associated with modeling the fast TQ in 3-D. There is no MHD activity or vertical motion of the plasma during the TQ stage. Rapid losses of thermal energy occur during this time, with the post-TQ temperature set by a balance between thermal conduction, impurity radiation, and Ohmic heating \citep{clauser2021iter}. The post-TQ result serves as the initial condition for the 3-D nonlinear CQ simulations described next. \autoref{tab:table} lists values of the post-TQ Lundquist number $S$ and the on-axis parallel electric field (normalized by the Connor-Hastie field $E_C$) for each case at the end of the TQ simulation ($t = \SI{0.1}{\milli\second}$). After the TQ, we decrease the thermal diffusivity by a factor of $10^3$ over the duration of the CQ, such that the plasma temperature is set primarily by the competition between Ohmic heating and impurity radiation. In the simulation without impurities (case A), and hence without radiative losses, the thermal diffusivity is kept constant at a large value $\chi_\perp \approx 10^5\SI{}{\meter \squared \per\second}$ after the initial TQ, causing the post-TQ temperature to remain low. Although Ohmic heating can cause reheating of the plasma in the absence of impurity radiation and limit RE generation, we do not model this effect in this case A, since we are interested in isolating effects that arise from the coupling of MHD activity with RE generation, as described in \blue{\S}\ref{sec:case_A}. 

MHD instabilities during flattop operation and the TQ can modify the plasma parameters and flatten the safety factor. However, we neglect these effects in the present simulations, which only comprise the CQ phase of the disruption. We further note that the simulations presented in \blue{\S}\ref{sec:results} are not meant to be fully predictive, but to guide our understanding of RE generation during SPARC disruptions, and to identify effects that arise from self-consistent MHD + RE coupling.


\begin{figure}[t!]
\includegraphics[page=15,width=0.48\textwidth]{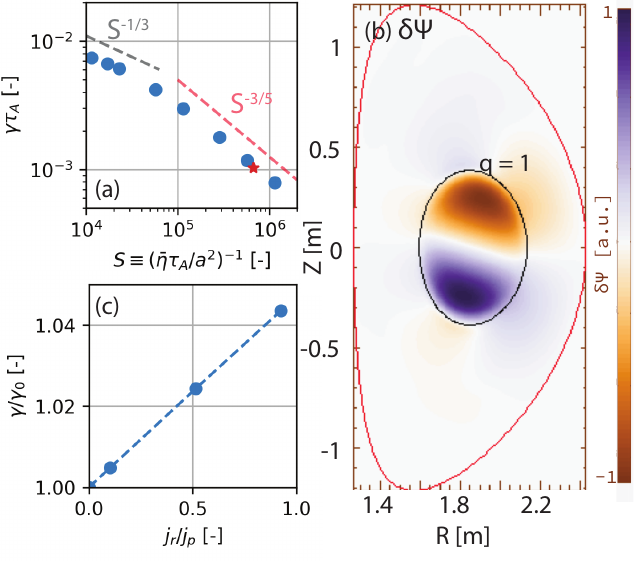}
\centering
\caption{(a)~Linear growth rate of the $m/n = 1/1$ mode as a function of the Lundquist number. Blue circles assume spatially-uniform resistivity, while the red star uses non-uniform Spitzer resistivity, and represents the actual conditions in the post-TQ plasma used for the nonlinear simulation shown in \autoref{fig:unmitigated_3d}. (b)~Linear Growth Rate of the $m/n = 1/1$ mode as a function of the RE current fraction.  Here, $\gamma_0 \approx \SI{1e-3}{}\tau_A^{-1}$ is the growth rate for $j_r/j_p = 0$ RE fraction. (c)~The perturbed poloidal flux $\delta\psi$ and obtained from a linear simulation of $n = 1$ modes in the post-TQ plasma. Here, the RE current is set to 0. Temperature-dependent Spitzer resistivity is used. The black contour represents the $q = 1$ flux surface.  
}
\label{fig:linear}
\end{figure}

\section{Results}

\label{sec:results}

\subsection{Idealized Disruption without Impurities}
\label{sec:case_A}

First, we report results from case A, which describes an idealized disruption without any impurities. The simulation is performed with a perfectly conducting first wall (M0 geometry; see \autoref{fig:geometry}\blue{a}) and 4 toroidal planes (resolving $n < 4$ modes).

The post-TQ conditions are similar to that reported previously in \blue{Ref.}~\citenum{datta2025coupled}. During the artificial TQ, the total thermal energy falls from $\SI{20}{\mega\joule}$ to about $\SI{50}{\kilo\joule}$ between $0 < t < \SI{0.05}{\milli\second}$, and remains mostly constant thereafter. The core temperature falls from $T_e \approx \SI{20}{\kilo\electronvolt}$ to about $T_e\approx 30-\SI{40}{\electronvolt}$. The post-TQ electric field is peaked-off axis (around $\sqrt{\psi_N} \approx 0.7$), with a maximum magnitude of $E_\parallel \approx \SI{70}{\volt\per\meter}$. The electric field, normalized by the Connor-Hastie critical field $E_C = e^3n_e \ln \Lambda / (4 \pi \epsilon_0^2 mc^2)$ \citep{connor1975relativistic}, ranges between $E_\parallel/E_C \approx 200\text{-}500$. The Lundquist number in the post-TQ plasma is about $S \equiv [(\eta/\mu_0)\tau_A/a^2]^{-1}\approx \SI{7e5}{}$. Here, $\eta$ is the flux-averaged resistivity calculated at the $q = 1$ surface, and $\tau_A =  R_0 / v_A = R_0 / (B_0/\sqrt{\mu_0 \rho}) \approx \SI{1.2e-7}{\second}$ is the Alfvén time calculated using the toroidal magnetic field $B_0 = \SI{12.2}{\tesla}$ and the volume-averaged ion density $\langle n_i \rangle \approx \langle \rho \rangle / M \approx \SI{3e20}{\per\cubic\meter}$.

\autoref{fig:unmitigated_3d}\blue{a} shows the evolution of the RE current $I_{RE}$ and the total plasma current $I_p$ (which comprises RE and Ohmic components) during the CQ. A final plateau current of $I_\text{f} \approx \SI{6.1}{\mega\ampere}$ is obtained around $\SI{6}{\milli\second}$, demonstrating roughly 70\% conversion of the pre-disruption Ohmic current into the RE current. The dashed lines in \autoref{fig:unmitigated_3d}\blue{a} represent the solution of an equivalent {2-D} axisymmetric simulation, without any MHD activity. We observe an enhancement of the RE current in the 3-D case when compared to the 2-D case. This can be seen more easily in \autoref{fig:unmitigated_3d}\blue{(d-e)}, which shows a roughly $\SI{0.12}{\mega\ampere}$ increase in the RE current between $\SI{0.92}{\milli\second} < t < \SI{0.97}{\milli\second}$, during the time of MHD activity. In the 2-D simulation, plateau formation occurs later, around \SI{7}{\milli\second} after the TQ onset. However, the final plateau current ($I_\text{f,2D} \approx \SI{5.9}{\mega\ampere}$) is similar, although slightly lower than in the 3-D case.

The enhancement in the RE generation rate coincides with MHD activity in the 3-D simulation. \autoref{fig:unmitigated_3d}\blue{b} shows the temporal evolution of the mode kinetic energy for different toroidal mode numbers $n < 4$. Periodic sawtooth-like  activity occurs, driven by a dominant $m,n = (1,1)$ internal mode. Higher-order $n$ modes are also observed, excited via nonlinear coupling. We calculate the average field line perturbation level using $|\delta B/ B_0| \approx \sqrt {W_\text{mag}(n) / W_\text{mag}(n=0)}$, where $W_\text{mag}(n)$ is the total magnetic energy for a given $n$ mode \citep{tinguely2021modeling,izzo2022runaway}. The maximum magnetic perturbation in this simulation is about $|\delta B/ B_0|\sim 10^{-3}$, as shown in \autoref{fig:unmitigated_3d}\blue{c}. 

For this simplified case, we perform a series of linear simulations to characterize the dominant $n = 1$ instability. In the linear simulations, we artificially vary the post-TQ Lundquist number by changing the plasma resistivity. These linear simulations assume spatially uniform resistivity and zero RE current. \autoref{fig:linear}\blue{a} shows  the linear growth rate $\gamma\tau_A$ as a function of the Lundquist number $S$. At low $S$, the growth rate scales as $\gamma \tau_A \sim S^{-1/3}$, consistent with the resistive kink scaling, while at high $S$, a $\gamma \tau_A \sim S^{-3/5}$ scaling is obtained, showing a transition to a tearing-type instability \citep{furth1963finite,coppi1976resistive,hastie1987stability}. The red star in \autoref{fig:linear}\blue{a} represents the linear result for the non-uniform temperature-dependent Spitzer resistivity case [$S \approx \SI{7e5}{}$] (that is, the initial condition for the 3-D nonlinear simulation). The observed growth rate agrees well with the $\sim S^{-3/5}$ tearing-type scaling \citep{furth1963finite,hastie1987stability}. \autoref{fig:linear}\blue{b} shows the perturbed poloidal flux $\delta \psi_p$ in this linear simulation. As expected, the dominant mode exhibits a $m = 1$ poloidal structure, enclosed within the $q = 1$ resonant surface. This linear simulation is then repeated by setting a temporally constant RE current density fraction $j_r/j_p$, similar to linear studies performed by various authors \citep{liu2021self,cai2015influence,sainterme2024resistive}. Here, $j_r$ is the RE current density, and $j_p$ is the total plasma current density. The linear growth rate for different values of $0 < j_r/j_p < 1$ is shown in \autoref{fig:linear}\blue{c}. The RE current causes a small increase in the linear growth rate, consistent with previously reported results \citep{cai2015influence,matsuyama2017reduced,liu2021self}.

 Having characterized the linear behavior of the $m,n = (1,1)$ mode, we return to the 3-D nonlinear results. \autoref{fig:unmitigated_3d}\blue{(f-g)} show the RE current density at $t = \SI{0.85}{\milli \second}$ and $t = \SI{0.94}{\milli \second}$ respectively, which is around the time of the first sawtooth. Initially, the RE current density exhibits off-axis peaking (\autoref{fig:unmitigated_3d}\blue{f}), similar to the result obtained in 2-D axisymmetric CQ simulations, due to an off-axis maximum in the electric field \citep{datta2025coupled}. At $t = \SI{0.94}{\milli\second}$, increased RE current density within the $q = 1$ surface is observed; furthermore, the RE current is localized within the $(1,1)$ magnetic island that forms during the pre-cursor to the sawtooth crash, as shown in \autoref{fig:unmitigated_3d}\blue{g}. Similar enhancement of the RE density by MHD activity has been reported previously in reduced MHD + RE fluid simulations performed in EXTREM \citep{matsuyama2017reduced}, where strong electric fields generated by the growth of MHD instabilities caused increased RE generation, primarily due to the exponential sensitivity of Dreicer generation to $E_\parallel$ \citep{dreicer1959electron,matsuyama2017reduced}. Indeed, when we repeat the 3-D simulation  without the Dreicer term, this enhancement is suppressed, as illustrated in \autoref{fig:unmitigated_3d}\blue{d}.


\begin{figure}[b!]
\includegraphics[page=8,width=0.48\textwidth]{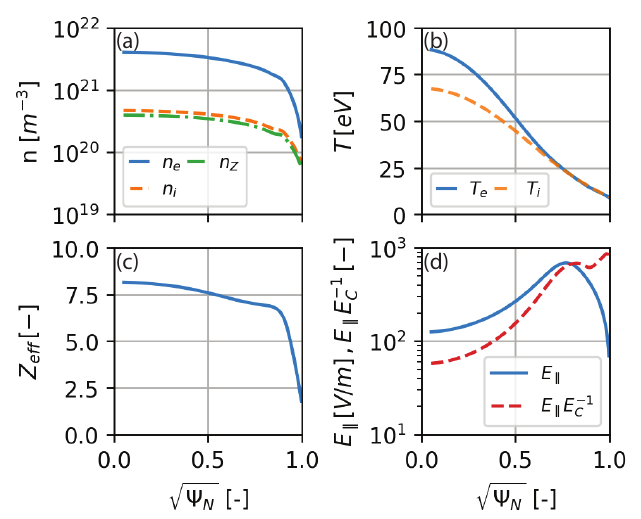}
\centering
\caption{ The flux-averaged post-TQ (a) electron $n_e$, deuterium $n_i$, and neon $n_Z$ densities, (b) electron and ion temperature, (c) effective ionization $Z_\text{eff}$, and (d) parallel electric field $E_\parallel$ at $t = \SI{0.1}{\milli\second}$ in case~B ($\SI{4.8e21}{}$ Ne atoms). The red dashed line is $E_\parallel$ normalized by $E_C$.
}
\label{fig:mitigated_TQ}
\end{figure}

\begin{figure*}[t!]
\includegraphics[page=16,width=0.99\textwidth]{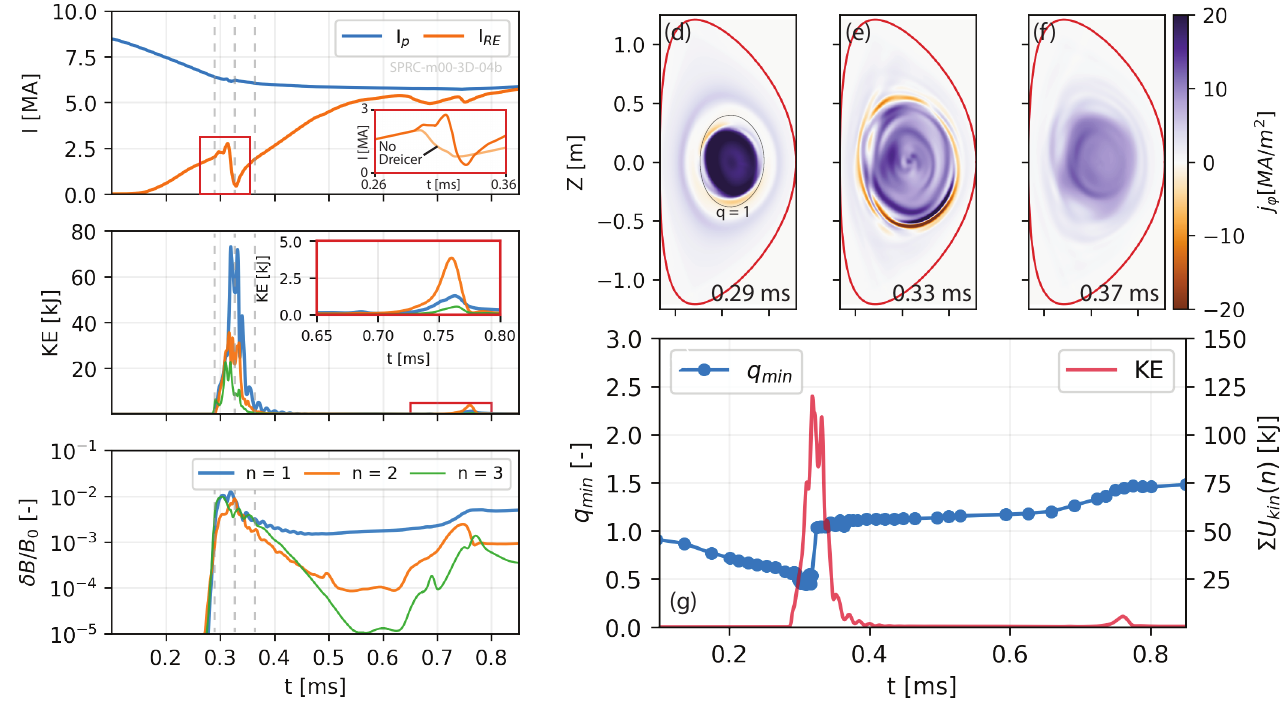}
\centering
\caption{ (a)~Evolution of plasma (blue) and runaway electron (orange) currents in the 3-D nonlinear simulation with $\SI{4.8e21}{}$ Ne atoms (case B). Inset: RE current with and without Dreicer generation. The Dreicer term is turned off at roughly 0.29~ms. (b)~Kinetic energy as a function of toroidal mode number $n$. (c)~Magnetic field line perturbation as a function of toroidal mode number $n$. (d-f)~Total plasma current density (at the $\phi = 0$ plane) at $t = \SI{0.29}{\milli \second}$, $t = \SI{0.33}{\milli \second}$, and $t = \SI{0.37}{\milli \second}$. (g)~Variation of minimum safety factor $q_{min}$ (blue, left) with time, together with the total kinetic energy $\Sigma_{n=1}^3 U_{kin} (n) $ of the $1 \leq n \leq 3$ modes (red, right). The markers represent the times at which the {M3D-C1} results were output. 
}
\label{fig:mitigated_CQ}
\end{figure*}

In the 3-D nonlinear simulation, the $m,n = (1,1)$ mode drives periodic sawtooth-like activity, characterized by a cyclical peaking and subsequent relaxation of the current profile. As observed in \autoref{fig:unmitigated_3d}\blue{(b-c)}, the kinetic energy and the magnetic perturbation of each successive sawtooth saturate at a lower value as the RE current increases. \autoref{fig:unmitigated_3d}\blue{h} shows the evolution of the square root of the poloidal flux $\sqrt{\psi_N}$ at the $q=1$ surface between $\SI{2}{\milli\second} < t < \SI{6}{\milli\second}$, together with the total kinetic energy of the $0<n<4$ modes. {Prior to each crash, the on-axis peaking of the current profile causes the central safety factor to fall further below unity. Consequently, the $q=1$ surface moves outwards radially, increasing the poloidal flux enclosed within it. During the crash, MHD activity redistributes current from the $q < 1$ region to the outer regions of the plasma, flattening the current profile, and causing $\psi_{N}(q=1)$ to fall. As seen in \autoref{fig:unmitigated_3d}\blue{h}, the flux enclosed by the $q = 1$ surface before each crash decreases with time, indicating reduced peaking of the current profile between successive crashes. When plateau formation occurs around \SI{6}{\milli\second}, the safety factor profile remains mostly invariant.} Notably, a $q = 1$ surface still exists in the plasma after plateau formation, which may cause a weaker crash at a later time. A similar suppression of nonlinear sawtooth activity driven by $1/1$ resistive kink modes was previously reported by \citeauthor{cai2015influence}. This effect can be explained in qualitative terms, by considering that the peaking of the current profile can be driven by both classical heating and RE avalanching \citep{eriksson2004current,cai2015influence}. These processes scale with the electric field $E_\parallel$, becoming weaker as $I_{RE}$ increases during the CQ. A comparison of our {M3D-C1} results with and without REs (not shown here) shows that in the absence of REs, the energy of the nonlinear modes remains much higher, although a decrease is also observed as the Ohmic current decays naturally during the CQ.

Lastly, despite MHD activity, RE losses are not observed in this simulation. Poincaré plots (not shown here) demonstrate that although MHD activity distorts the flux surfaces, forming a $(1,1)$ magnetic island and a displaced core within the $q < 1$ region, the flux surfaces remain intact after the crash, confining the REs. 

\begin{figure*}[t!]
\includegraphics[page=17,width=0.8\textwidth]{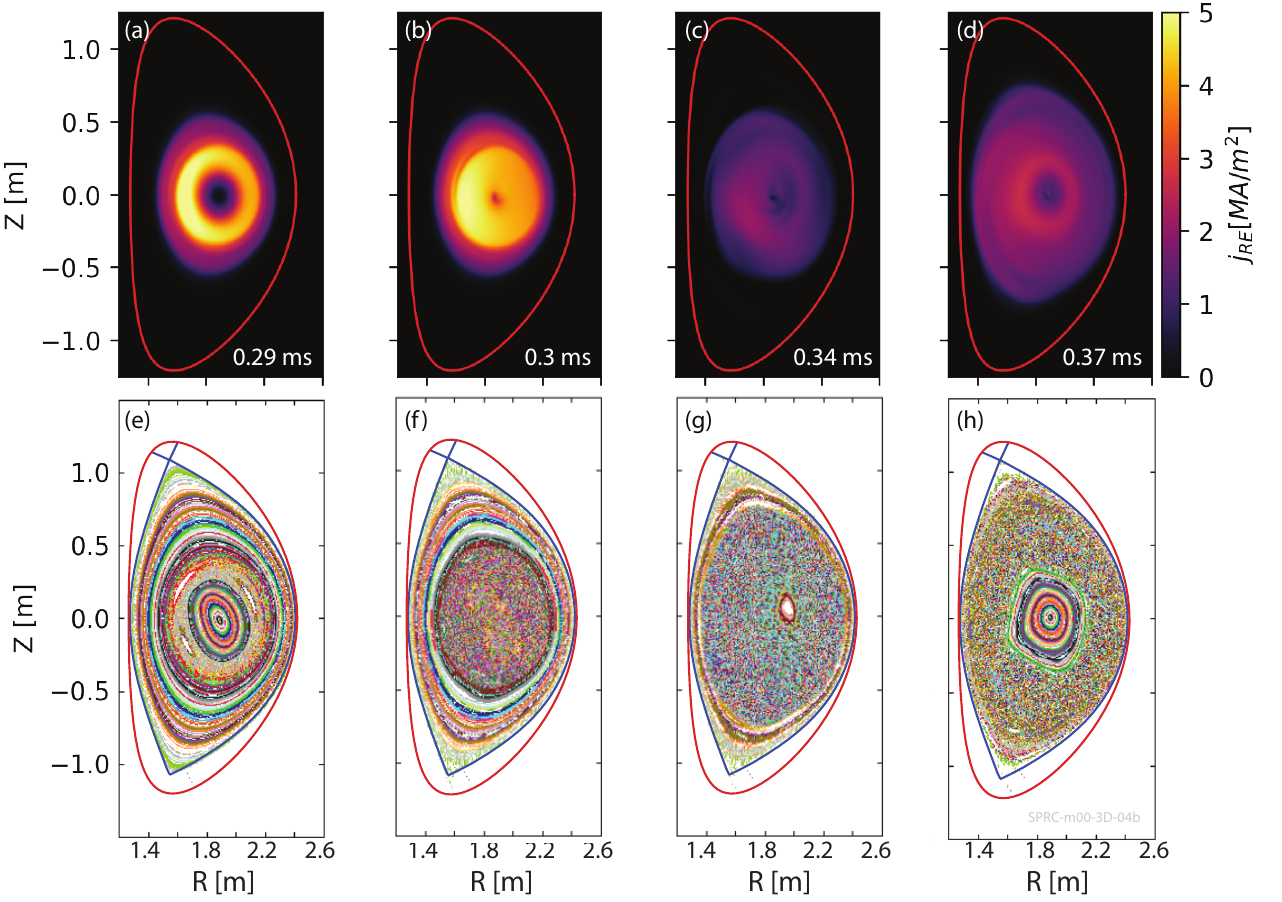}
\centering
\caption{ (a-d) RE current density (at the $\phi = 0$ plane) at different times during the first MHD event in the 3-D nonlinear simulation with $\SI{4.8e21}{}$ Ne atoms (Case B). Stocastic field lines enable radial transport of the RE current density.  (d-h)~Poincaré plots showing the magnetic field topology at the same times. The red contour marks the simulation domain, while the blue line is the flux contour at the plasma boundary. 
}
\label{fig:poincare_mitigated}
\end{figure*}

\subsection{Disruptions with Neon Injection}



Next, we describe case B, where we inject roughly $\SI{4.8e21}{}$ Ne atoms at $t = \SI{0}{\milli\second}$. Again, this simulation is performed with a perfectly conducting first wall (M0 geometry, see \autoref{fig:geometry}\blue{a}) and 4 toroidal planes.
\autoref{fig:mitigated_TQ}\blue{a} shows the flux-averaged post-TQ profiles of the total electron density $n_e$, (main) ion density $n_i$, and neon density $n_Z$, while \autoref{fig:mitigated_TQ}\blue{(b-c)} shows the flux-averaged electron and ion temperatures, and the effective ionization $Z_\text{eff} \equiv \sum n_j Z_j^2 / n_{\mathrm{e}}$ at $t = \SI{0.1}{\milli\second}$. The value of $Z_\text{eff}$ peaks at the core, where the ion temperature is highest, contributing the large electron density in this region. \autoref{fig:mitigated_TQ}\blue{d} shows the flux-averaged parallel electric field $E_\parallel$. Similar to the previous case,  $E_\parallel$ is peaked off-axis around $\sqrt{\psi}_N\approx 0.7$. The magnitude of $E_\parallel$ varies between $120-\SI{700}{\volt\per\meter}$, significantly higher than in the previous case, primarily because the larger $Z_\text{eff}$ contributes to a higher (Spitzer) resistivity $\eta_\parallel \sim Z_\text{eff}T_e^{-3/2}$. The post-TQ plasma exhibits a Lundquist number of roughly $S \approx \SI{5e4}{}$. The electric field normalized by the critical field $E_C$ is about $E_\parallel/E_C \approx 60$ near the core where the electron density high and increases to about $E_\parallel/E_C \approx 800$ further away from the magnetic axis (\autoref{fig:mitigated_TQ}\blue{d}).

\blue{\autoref{fig:mitigated_CQ}a} shows the temporal evolution of the total plasma current $I_p$ and the RE current $I_{RE}$ during the CQ. The RE current grows initially, but between $\SI{0.31}{\milli\second}< t <\SI{0.33}{\milli\second}$, the RE current decreases from about $I_{RE} \approx \SI{2.76}{\mega\ampere}$ to roughly $\SI{0.43}{\mega\ampere}$. Thereafter, the RE current increases again, reaching a value of about $I_{RE} \approx \SI{5.5}{\mega \ampere}$ around \SI{0.6}{\milli\second}. Smaller, intermittent RE losses occur between $\SI{0.65}{\milli\second}< t <\SI{0.8}{\milli\second}$, before eventual plateau formation, with a final RE current of roughly $I_f \approx \SI{5.7}{\mega\ampere}$. This value agrees well with that in NIMROD + DREAM simulations, performed with a similar quantity of Ne \citep{tinguely2021modeling}. 

Right before the first loss event around $t \approx \SI{0.3}{\milli\second}$, we observe a rapid increase in the RE current caused by the Dreicer term, similar to that in the previous simulation with no impurities. During this period, the RE current increases by about $\SI{0.7}{\mega\ampere}$, before eventually falling during the loss event. This is shown in the inset of \blue{\autoref{fig:mitigated_CQ}a}, which compares the RE current evolution during the MHD event with and without the Dreicer term.

RE losses coincide with strong MHD activity, as seen in \blue{\autoref{fig:mitigated_CQ}b}, which shows the kinetic energy associated with different toroidal $n$ modes. The initial onset of MHD activity is dominated by $n = 3$ toroidal modes, but the $n = 1$ modes eventually become dominant. The $n = 1$ mode kinetic energy saturates at roughly $\SI{75}{\kilo\joule}$ around $t \approx \SI{0.32}{\milli\second}$. \blue{\autoref{fig:mitigated_CQ}(d-f)} shows the plasma current density before ($t = \SI{0.29}{\milli\second}$), during ($t = \SI{0.33}{\milli\second}$), and after ($t = \SI{0.37}{\milli\second}$) the MHD event. Initially, a contraction of the current density profile occurs, which is accompanied by a decrease in the minimum safety factor. \autoref{fig:mitigated_CQ}\blue{g} shows the variation of $q_{min}$ with time, together with the total kinetic energy of the $1 \leq n \leq 3$ modes. The value of $q_{min}$ falls to about $q_{min} \approx 0.5$ at $t = \SI{0.29}{\milli\second}$. The initial onset of MHD instabilities occurs near the $q = 1$ surface (see \autoref{fig:mitigated_CQ}\blue{d}), and by $t = \SI{0.33}{\milli\second}$, the current perturbations have penetrated into the core (\blue{\autoref{fig:mitigated_CQ}e}). Lastly, while the current profile is highly peaked at \SI{0.29}{\milli\second}, an inspection of the current density after the MHD event at $t = \SI{0.37}{\milli\second}$ demonstrates a flatter current profile (and thus, lower internal inductance), as seen in \blue{\autoref{fig:mitigated_CQ}f}. A small $I_p$ spike accompanies the MHD event, as seen in \autoref{fig:mitigated_CQ}\blue{a}. After MHD activity, the safety factor becomes $q > 1$, as seen in \autoref{fig:mitigated_CQ}\blue{g}. 

The average field line perturbation level is shown in \autoref{fig:mitigated_CQ}\blue{c}. This value peaks at about $\delta B / B_0 \sim 10^{-2}$, which is an order of magnitude larger than in the previous simulation. We visualize the RE current density and magnetic field topology between $\SI{0.29}{\milli\second}< t <\SI{0.37}{\milli\second}$ in \autoref{fig:poincare_mitigated}. Initially, the core becomes stochastic, while a band of intact flux surfaces separates the stochastic core from the first wall, as shown in \autoref{fig:poincare_mitigated}\blue{f}. During this time, no RE losses occur. Instead, fast transport within the core simply homogenizes the RE current in this region (\autoref{fig:poincare_mitigated}\blue{b}), which is initially peaked off-axis at \SI{0.29}{\milli\second} (\autoref{fig:poincare_mitigated}\blue{a}). Eventually, the field lines in the outer region become stochastic, which enables the transport of REs towards the first wall, as seen in \autoref{fig:poincare_mitigated}\blue{(c,g)}. During this time, the RE current decreases as the REs are transported across the last closed flux surface (LCFS) into the region of open field lines. Our simulation demonstrates a $\Delta I_{RE} \approx \SI{2.3}{\mega\ampere}$ decrease in the RE current over a period of roughly $\Delta t \approx \SI{0.02}{\milli\second}$, as shown in \autoref{fig:mitigated_CQ}. During the RE loss, $I_p$ does not change significantly. Instead, as noted earlier, we see a small $I_p$ spike, and a conversion of the RE current into Ohmic current occurs. The post-crash $\delta B / B_0$ decreases to about $B / B_0 \sim 10^{-3}$ after $t > \SI{0.32}{\milli\second}$, set primarily by the $n=1$ perturbation. During this period, the RE current increases again. At \SI{0.37}{\milli\second}, rehealed flux surfaces are observed in the core, whereas the outer region is still stochastic (\autoref{fig:poincare_mitigated}\blue{h}). Intact flux surfaces enable the subsequent growth of the RE population via avalanching, despite the preceding loss event. In this simulation, the onset of flux surface re-formation begins when $q_{min}$ rises above unity.

\begin{figure}[t!]
\includegraphics[page=18,width=0.49\textwidth]{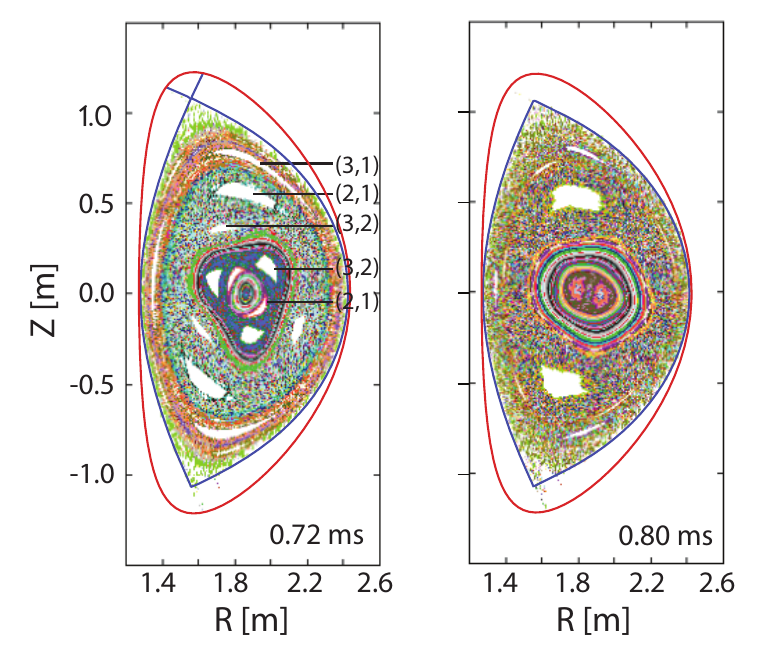}
\centering
\caption{Poincaré plots showing the magnetic field topology at \SI{0.72}{\milli\second} and \SI{0.8}{\milli\second} respectively (Case B).
}
\label{fig:poincare_islands}
\end{figure}

Similar qualitative behavior is observed during the second MHD event that occurs around $t = \SI{0.75}{\milli\second}$.  The field lines become stochastic and enable RE transport, whereas the re-formation of flux surfaces limits the level of RE losses. Here, the $n = 2$ toroidal modes exhibit the highest kinetic energy, saturating at roughly \SI{5}{\kilo\joule}, as shown in the inset of \autoref{fig:mitigated_CQ}\blue{b}. An inspection of $q(\psi_N)$ shows that the $q$-profile is non-monotonic, with $q_{min} \approx 1.2$ prior to the onset of the MHD activity, as illustrated in \autoref{fig:mitigated_CQ}\blue{g}. $m,n = (2,1), (3,2), \text{and } (3,1)$ magnetic islands appear within a sea of magnetic stochasticity (see \autoref{fig:poincare_islands}). As $q_{min}$ increases to $q_{min} \gtrsim 1.5$, $n = 1$ modes become the dominant perturbation. After the nonlinear saturation of MHD activity, the RE current grows and becomes equal to the total plasma current $I_p$ around $t \approx \SI{0.85}{\milli\second}$.

\begin{figure*}[t!]
\includegraphics[page=20,width=0.99\textwidth]{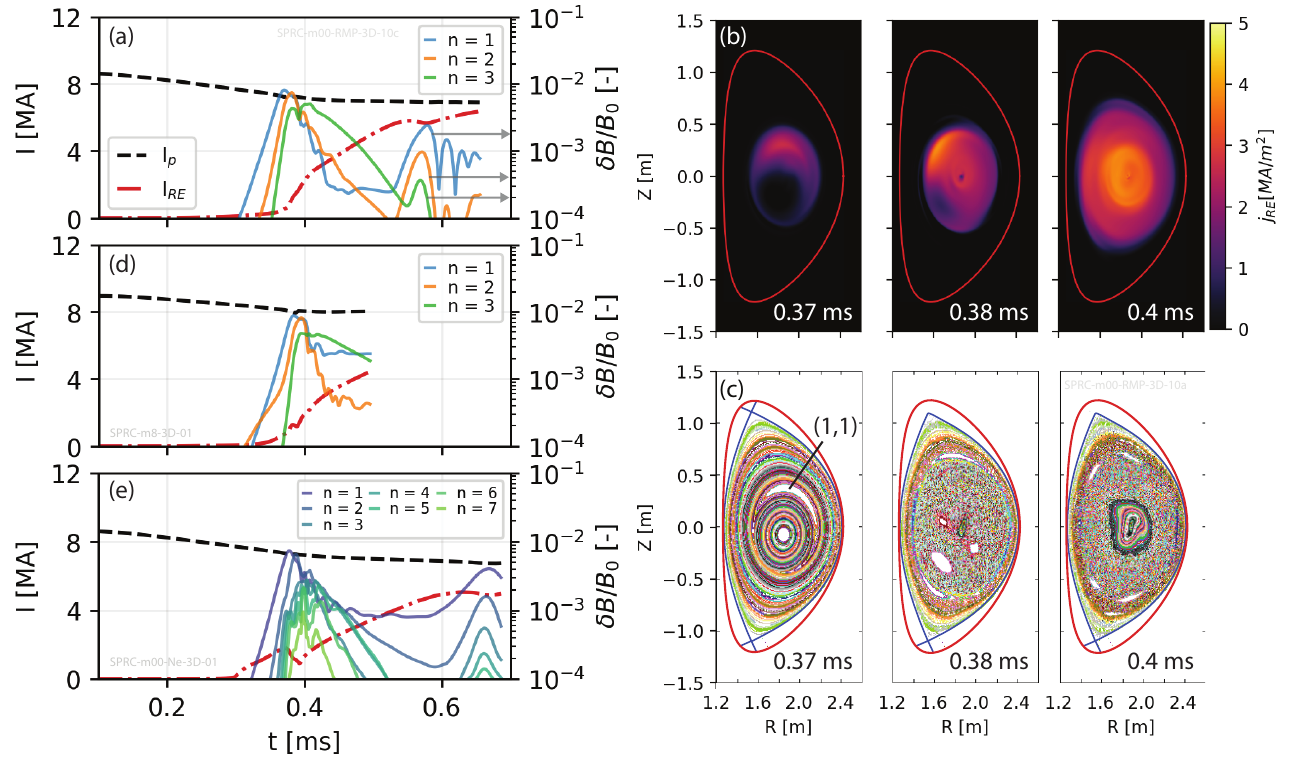}
\centering
\caption{(a)~[Left axis] Evolution of the plasma current (black, dashed) and RE current (red, dot-dashed) in a disruption with $\SI{2e21}{}$ Ne atoms (Case C). [Right axis] Magnetic perturbation level as a function of time. (b)~RE current density and (c) Poincaré plots of the magnetic topology at $t = (0.37,0.38,0.4)$ \SI{}{\milli\second}. (d)~Case C with the double-walled VV geometry (M1) and 4 toroidal planes. (e)~Case C with 8 toroidal planes (M0 geometry).
}
\label{fig:case_C}
\end{figure*}

\textit{Lower Neon Injection –} We repeat the previous simulation with a lower level of Ne (case C, $\SI{2e21}{}$ Ne atoms). \autoref{fig:case_C}\blue{a} shows the evolution of the plasma and RE currents and the magnetic perturbation level $\delta B/B_0$ of $n < 4$ toroidal modes during the CQ. Here, the RE current increases to roughly \SI{6}{\mega\ampere} by $t \approx \SI{0.65}{\milli\second}$. Similar to the previous case, a large MHD event occurs around $t \approx \SI{0.4}{\milli\second}$, followed by a smaller-amplitude event around $t \approx \SI{0.58}{\milli\second}$. Around $t \approx \SI{0.4}{\milli\second}$, the MHD activity saturates at roughly $\sum_{n=1}^3 U_{kin}(n) \approx \SI{50}{\kilo\joule}$, and the magnetic perturbation level is $\delta B/B_0 \approx 10^{-2}$. Although stochastic field lines are generated during the first crash, a decrease in the RE current is not observed during this period. On the other hand, the RE current decreases by $\Delta I_{RE} \approx \SI{0.2}{\mega\ampere}$ during the second MHD event. Similar to the previous case, the onset of field line stochasticity initially homogenizes the RE density within the core and transports REs across the poloidal plane. However, the rapid re-formation of good flux surfaces in the core confines REs and limits the loss of RE current. These dynamics are illustrated in \autoref{fig:case_C}\blue{(c-d)}. Initially, a $(1,1)$ magnetic island forms near the $q = 1$ flux surface. At this time ($t \approx \SI{0.37}{\milli\second}$), the RE current density is peaked off-axis and is further localized within the magnetic island. By $t \approx \SI{0.38}{\milli\second}$, the flux surfaces have broken and become stochastic, enabling fast transport of REs. However, stochasticity is short-lived, and by $t \approx \SI{0.4}{\milli\second}$, flux surfaces have re-healed within the core.

As noted earlier, this simulation assumes a perfectly conducting boundary at the first wall, which has a stabilizing effect on MHD instabilities, especially on higher $m/n$ modes that appear closer to the first wall. We repeat case C with the double-walled VV geometry, as described earlier in \blue{\S}\ref{sec:methods} (see \autoref{fig:geometry}\blue{c}). This simulation was run until $\SI{0.5}{\milli\second}$, capturing the first MHD event. The results are shown in \autoref{fig:case_C}\blue{d}. The evolution of the RE current and the MHD activity is largely similar to the simulation performed with the perfectly conducting first wall (\autoref{fig:case_C}\blue{a}). However, we now observe a small RE loss event around $\SI{0.4}{\milli\second}$. The total kinetic energy of the $n<4$ toroidal modes saturates at roughly $\SI{60}{\kilo\joule}$ in this simulation, while $n = 2$ modes appears to dominate the very early time growth of MHD activity. This is consistent with a relaxation of the perfectly conducting condition near the plasma boundary. Nevertheless, the differences seen in the more realistic geometry do not constitute a significant departure from the simpler perfectly conducting case. This is likely because the dynamics modeled here evolve on time scales much faster than the resistive wall time on SPARC. 

The simulations described previously were performed with 4 toroidal planes. While this toroidal resolution is sufficient to resolve the dominant modes in our simulation, previous work indicates that sub-dominant higher $n$ modes can contribute to field line stochasticity \citep{igochine2006stochastic,liu2021self}. We repeat case C (M0 geometry, $\SI{2e21}{}$ Ne atoms) with 8 toroidal planes to quantify the effect of higher $n$ modes on the RE evolution. \autoref{fig:case_C}\blue{e} shows the results, which exhibit some key differences compared to the 4-plane simulation. Firstly, an earlier initial rise in the RE current occurs in this simulation. Secondly, the RE current decreases by roughly $\Delta I_{RE} \approx \SI{1}{\mega\ampere}$ during the first MHD event. The contribution of the sub-dominant higher-$n$ modes increases the total kinetic energy, which saturates at about $\sum_{n=1}^7 U_{kin}(n) \approx \SI{75}{\kilo\joule}$. MHD instabilities grow again around $t \approx \SI{0.65}{\milli\second}$, and the RE current decreases by about $\Delta I_{RE} \approx\SI{0.2}{\mega\ampere}$. RE loss during the second MHD event is similar in the 8- and 4-plane simulations. 


\subsection{Combined Deuterium and Neon Injection}

\begin{figure}[t!]
\includegraphics[page=26,width=0.48\textwidth]{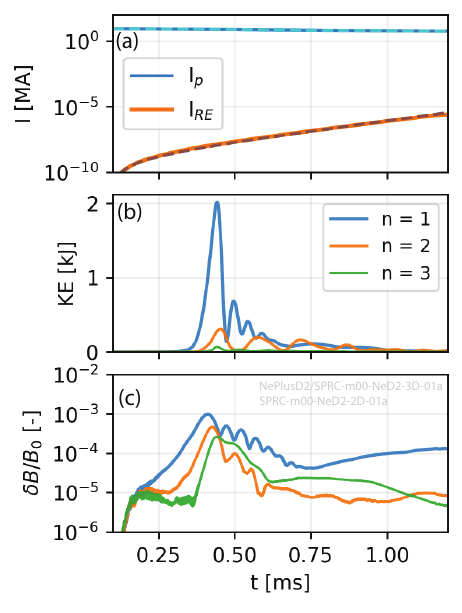}
\centering
\caption{(a)~Plasma (blue) and RE (orange) currents in a 3-D simulation with combined Ne + $\text{D}_2$ injection (case D). Dashed purple and red lines represent $I_p$ and $I_{RE}$ in an equivalent 2-D axisymmetric simulation. (b-c) Mode kinetic energy, and $\delta B/B_0$ in the 3-D simulation (case D). This simulation was performed with the M0 geometry.}
\label{fig:case_D}
\end{figure}

As described in \blue{\S}\ref{sec:intro}, the SPARC baseline MGI case will comprise simultaneous injection of $\text{D}_2$ + Ne to mitigate thermal and RE loads \citep{sweeney2020mhd,ekmark2025runaway}. \autoref{fig:case_D}\blue{a} shows the temporal evolution of the RE current for case D, which comprises $\SI{2e21}{}$ Ne atoms and $\SI{1.8e22}{}$ $\text{D}_2$ molecules. The solid lines show the result for a 3-D simulation, performed with the perfectly conducting first wall geometry (M0) and 4 toroidal planes, and the dashed lines show the results for an equivalent 2-D axisymmetric simulation. As expected, the RE current grows at a significantly slower rate with combined $\text{D}_2$ + Ne injection when compared to the Ne-only cases described previously (\autoref{fig:case_C}). After injection, the on-axis deuterium and neon densities are about $n_{D,0} \approx \SI{6e21}{\per\cubic\meter}$ and $n_{Ne,0} \approx \SI{1.7e20}{\per\cubic\meter}$ respectively, and the post-TQ parallel electric field is roughly $E_\parallel/E_C \approx 7$ on-axis at $t = \SI{0.1}{\milli\second}$. Furthermore, the electric field normalized by the Dreicer field \citep{dreicer1959electron} is $E_\parallel/E_D < \SI{7e-3}{}$, which effectively suppresses Dreicer generation in the $\text{D}_2$ + Ne case.

The 3-D simulation was run for about \SI{1.2}{\milli\second}, capturing MHD activity that occurs around $t \approx \SI{0.4}{\milli\second}$. \autoref{fig:case_D}\blue{b-c} shows the evolution of the mode KE and magnetic perturbation level $\delta B/B_0$.  The MHD activity is dominated by $n = 1$ toroidal modes. The total kinetic energy of the instabilities saturates around $\sum_{n=1}^3 U_{kin}(n) \approx \SI{2}{\kilo\joule}$, and the magnetic perturbation level is  $\delta B/B_0 < 10^{-3}$. No significant changes to the RE current occur due to MHD activity in this simulation, and the RE current is similar to that in the 2-D simulation, as shown in \autoref{fig:case_D}\blue{a}. Magnetic field lines remain intact during MHD activity in this simulation.
 
\begin{figure*}[t!]
\includegraphics[page=27,width=0.99\textwidth]{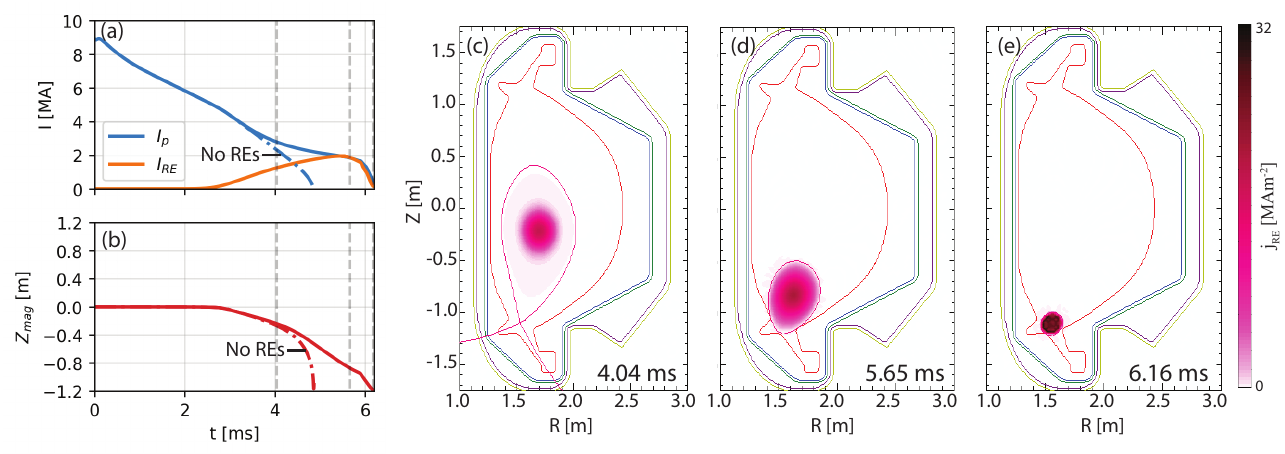}
\centering
\caption{(a)~Plasma (blue) and RE (orange) currents in a 2-D axisymmetric simulation with combined Ne + $\text{D}_2$ injection (case D), performed using the double-walled VV geometry (M1). The dashed blue line shows $I_p(t)$ in a simulation where REs are artificially suppressed. (b) The vertical position of the magnetic axis as a function of time. Dashed line shows $Z_{mag}$ from a simulation where the RE current is set to 0. (c-e) Evolution of the RE current density and the LCFS during the cold VDE at $t = (4.04,\, 5.65 \, \text{ and } \SI{6.16}{\milli\second})$ [vertical dashed lines in (a-c)].}
\label{fig:case_D_VDE}
\end{figure*}

\textit{2-D Cold VDE Simulation – } When the plasma current $I_p$ falls below a critical value, we expect a "cold" VDE to be triggered, which refers to a post-TQ displacement of the plasma column \citep{boozer2019halo,kiramov2022bifurcation}. This is in contrast to a "hot" VDE, where a loss of vertical control results in displacement, which is then followed by the TQ. In the simulation with the M0 geometry shown in \autoref{fig:case_D}, the use of time-invariant boundary conditions on the first wall precludes modeling of cold VDE dynamics. To capture the cold VDE, we perform a 2-D axisymmetric simulation of case D with the double-walled VV geometry (M1, see \autoref{fig:geometry}\blue{c}). \autoref{fig:case_D_VDE}\blue{a} shows the evolution of the plasma and RE currents in this simulation, while \autoref{fig:case_D_VDE}\blue{b} shows the variation in the vertical position of the magnetic axis $Z_{mag}$ with time. Vertical motion begins around $t \approx \SI{3}{\milli\second}$ when the plasma current is roughly $I_p \approx \SI{5}{\mega \ampere}$ \citep{clauser2024modeling}. Between $\SI{3}{\milli\second}< t < \SI{5.65}{\milli\second}$, the magnetic axis is displaced in the $-Z$ direction by about \SI{0.85}{\meter}, and the plasma moves downwards towards the lower divertor. This is seen in \autoref{fig:case_D_VDE}\blue{(d-f)}, which shows the RE current density and the LCFS at different times during the VDE. The RE current increases during this period, reaching a maximum value of $I_{RE} \approx \SI{2}{\mega \ampere}$ around $t \approx \SI{5.65}{\milli\second}$. At this time, $I_p \approx I_{RE}$ and all of the current is carried by the REs. Thereafter, the LCFS contacts the first wall (\autoref{fig:case_D_VDE}\blue{e}), and $I_{RE}$ decreases due to scrape-off until the RE beam is eventually terminated around $t \approx \SI{6.2}{\milli\second}$. Unlike the previous cases, where RE losses are accompanied by a conversion to Ohmic current, such conversion does not occur here; instead, the total plasma current $I_p$ continues to decrease during VDE.


The dashed lines in \autoref{fig:case_D_VDE}\blue{(a-c)} show the evolution of $I_{p}$ and $Z_{mag}$ when REs are artificially suppressed in the simulation. A faster vertical motion $\partial_t Z_{mag}$ of the magnetic axis and earlier termination of the plasma current (around $t \approx \SI{4.8}{\milli\second}$) is observed in the simulation without REs. As seen in \autoref{fig:case_D_VDE}\blue{a}, REs decrease the rate of decay of the plasma current $\partial_t I_p(t)$; this is expected since a non-negligible RE current density $j_{RE}$ decreases $E_\parallel = \eta_\parallel(j_p - j_{RE})$, thus reducing $\partial_t I_p \propto E_\parallel$. The plasma displacement during a cold VDE is typically related to the plasma current $I_p(t)$ \citep{boozer2019halo,clauser2021iter,kiramov2022bifurcation}, and the slower vertical motion observed in \autoref{fig:case_D_VDE} is consistent with a slower plasma current decay rate. M3D-C1 simulations of ITER disruptions (without REs) have previously shown a "freezing" of the vertical motion of the plasma with a freezing of the CQ \citep{clauser2021iter}. Similarly, a slowdown of the vertical motion of the plasma due to RE generation was also reported in some JOREK "hot" VDE simulations of ITER disruptions \citep{wang2024effect}.


\section{Discussion}
\label{sec:discussion}


The key results from \blue{\S}\ref{sec:results} can be summarized as follows:

(1) Electric fields induced by MHD instabilities can enhance the RE current. This occurs because of the exponential sensitivity of Dreicer generation to the electric field. This enhancement was observed in the simplified "sawtoothing" disruption plasma (Case A; see \autoref{fig:unmitigated_3d}) where the field lines remained intact, as well as in the simulations with Ne impurities, where rapid generation was observed during the initial stages of MHD instability growth (\autoref{fig:mitigated_CQ}\blue{a}). 

(2) An increasing RE current during the CQ can decrease the nonlinear amplitude of the $m,n = (1,1)$ instability that drives cyclical sawteething activity, as seen in \autoref{fig:unmitigated_3d}. This is consistent with reduced peaking of the current profile between successive crashes.

(3) Significant losses of RE occur when MHD activity generates stochastic field lines, as observed in simulations with Ne injection (Cases B-C; see \autoref{fig:mitigated_CQ} and \autoref{fig:case_D}). However, a reformation of intact flux surfaces in the core can lead to continued RE growth after the initial loss event (\autoref{fig:poincare_mitigated}). This indicates that for RE mitigation with stochastic magnetic fields, it is important to prevent re-healing of flux surfaces.

(4) Finally, while relatively large RE plateau currents are produced by Ne-only injection, combined Ne + $\text{D}_2$ injection results in a lower RE current in SPARC (see \autoref{fig:case_D_VDE}), consistent with previous simulations \citep{ekmark2025runaway} and experiments on other machines. Here, a "cold" VDE may be triggered, causing termination of the RE beam. REs were observed to slow down the vertical motion of the plasma, leading to later  termination.

The rapid increase in the RE current via Dreicer generation was first noted by \citeauthor{matsuyama2017reduced} in reduced MHD simulations of the $m,n = (1,1)$ resistive kink mode performed using EXTREM \citep{matsuyama2017reduced} and was further hypothesized in MGI-driven TQ simulations of NSTX performed using {M3D-C1} (but without a RE fluid model), where strong spatially localized spikes in the parallel electric field were observed during periods of MHD activity \citep{ferraro20183d}. Our results are largely in agreement with the reduced MHD results reported by \citeauthor{matsuyama2017reduced}. In case A, between $\SI{0.9}{\milli\second} < t < \SI{0.5}{\milli\second}$, when the enhancement in the RE current is observed (see \autoref{fig:unmitigated_3d}), $E_\parallel/E_D$ increases from $E_\parallel/E_D \approx 0.2$ to $E_\parallel/E_D \approx 0.26$ at the $q = 1$ surface. The small change in $E_\parallel/E_D$ increases the Dreicer rate significantly, from about $\SI{6.5e19}{\per \cubic \meter \per \second}$ to $\SI{1.35e21}{\per \cubic \meter \per \second}$ at this location. Furthermore, the Dreicer-driven enhancement is only observed at the first sawtooth, when the RE density is low enough that the avalanching rate does not fully dominate RE generation, consistent with EXTREM results \citep{matsuyama2017reduced}. Similar enhancements are also observed in the Ne-only MGI simulations, consistent with a rise in the electric field during the growth of MHD instabilities, as noted earlier in \blue{\S}\ref{sec:results}.  In the disruption with combined $\text{D}_2$ + Ne injection (Case D), however, this effect is not observed. Here, $E_\parallel/E_D$ is small enough ($< 0.01$) that Dreicer generation remains negligible despite the increased electric field. Changes in $E_\parallel$ can also modify the avalanching rate; however, these changes are less pronounced given the linear dependence of the avalanching rate on $E_\parallel$ \citep{rosenbluth1997theory}.


The suppression of sawteeth oscillations, observed in case A (see \autoref{fig:unmitigated_3d}), is also consistent with previous MHD + RE simulations of $(1,1)$ resistive kinks, performed using the code M3D \cite{cai2015influence}. Unlike our simulations, where the RE current grows continuously during the CQ, the M3D simulations assume a constant RE current and demonstrate that the post-crash safety factor remains flattened at $q \gtrsim 1$, preventing further oscillations when the RE current is the only current carrier \citep{cai2015influence}. Our simulations are consistent with this result, showing reduced peaking with increasing RE current and a flattened, largely invariant $q$-profile at the RE plateau, as seen in \autoref{fig:unmitigated_3d}\blue{h}. 

The loss of RE electrons in stochastic magnetic fields, as observed in cases B-C, is a well-known result, seen in various MHD + RE simulations \citep{liu2021self,zhao2024simulation,bandaru2021magnetohydrodynamic}. As expected, the level of RE losses depends strongly on the spatial distribution and longevity of stochastic fields. In both cases with Ne-only injection (Cases B-C), intact flux surfaces in the outer region only lead to an initial homogenization of RE density within the stochastic core (see \autoref{fig:poincare_mitigated} and \autoref{fig:case_C}). RE losses only occur when the outer flux surfaces break; however, the re-healing of flux surfaces within the core limits RE losses. The re-healing of flux surfaces and its significance for RE confinement have been emphasized previously, where a similar reformation of flux surfaces in the core was observed in MGI-driven disruption simulations of SPARC performed in NIMROD \citep{izzo2022runaway,izzo2025disruption}. These NIMROD simulations included REMC fields, which seed and sustain magnetic stochasticity; however, a re-healing of flux surfaces is observed when $q_{min} >2$ with the REMC \citep{izzo2025disruption}. The {M3D-C1} simulations presented here focus on MGI, but future work will explore coupling with the REMC for RE mitigation on SPARC.




Lastly, the slowest RE generation rate is obtained with combined $\text{D}_2$ + Ne injection (see \autoref{fig:case_D}), as expected from previous simulation and experimental results. Here, both RE and MHD dynamics are slowed down significantly  by the higher plasma density, which decreases both $E_\parallel/E_C$ and the Alfvén velocity $v_A$. 3-D MHD instabilities do not appear to significantly affect the RE evolution, likely due to the comparatively lower amplitudes of the MHD modes in this simulation (see \autoref{fig:case_D}). Therefore, our result is qualitatively similar to DREAM simulations of SPARC disruptions reported by \citeauthor{ekmark2025runaway}, which do not directly model MHD activity, and also demonstrate reduced RE currents with combined $\text{D}_2$ + Ne MGI \citep{ekmark2025runaway}. In SPARC, the goal of deuterium injection is to raise the electron density, rather than to recombine the plasma, as is done to benignly terminate a pre-existing RE beam by triggering external MHD instabilities \citep{paz2021novel}. Previous work indicates that the level of recombination due to $\text{D}_2$ injection is expected to be small in SPARC \citep{hollmann2023trends}. 


The DREAM results \citep{ekmark2025runaway} also exclude the vertical motion of the plasma during the disruption, although recent DREAM simulations of ITER disruptions have incorporated this effect in a reduced manner, by assuming that REs outside an instantaneous LCFS are scraped-off and lost to the wall \citep{vallhagen2025reduced}. As shown in \autoref{fig:case_D_VDE}, the decay of the plasma current below a critical value (here, about 5~MA) triggers a cold VDE, which eventually terminates the RE beam via scrape-off near the lower divertor region, precluding the formation of a steady plateau. This result indicates that the SPARC baseline MGI case may cause REs to strike the outer or inner off-midplane limiters, and that some contact with the divertor is possible, although 3-D instabilities (not modeled in the VDE simulation) may potentially limit this contact. REs also modify the dynamics of the VDE, further highlighting the importance of a self-consistent treatment of RE and MHD physics.

As noted earlier in \blue{\S}\ref{sec:methods}, the {M3D-C1} simulations presented in the paper use simplifying assumptions to manage the numerical complexity and computational expense of these calculations. For instance, the cold VDE was modeled in a 2-D axisymmetric simulation of combined $\text{D}_2$ + Ne injection, and 3-D MHD instabilities were only considered in a separate 3-D simulation without vertical motion of the plasma, given the computational expense of simulating 3-D non-axisymmetric VDEs. During the VDE, changes in the $q$-profile can trigger MHD instabilities, which can accelerate RE losses if they produce sufficient field line stochasticity. In the VDE simulation (\autoref{fig:case_D_VDE}), the edge safety factor  $q_{95}$ falls below 2 around $t \approx \SI{5.6}{\milli\second}$, around the time the LCFS contacts the first wall. This will likely trigger external MHD modes \citep{turnbull2016external,paz2019kink} that modify the termination of the RE beam. The coupling of 3-D MHD instabilities, RE generation, and scrape-off during VDEs will be the topic of a separate study. 



 We also ignore MHD activity during the TQ in the present simulations and  assume a rather idealized distribution of impurities at the start of the TQ. Resistive effects were found to be small in the 3-D simulations; however, they could become important for disruptions that evolve over longer time scales, which may also depend on MHD activity during the pre-TQ and TQ phases. In addition, hot-tail generation, which is excluded from the present simulations, is expected to dominate primary generation during the TQ \citep{smith2008hot}, whereas Dreicer and activated sources are only important when the temperature becomes low enough for the electric field to exceed the critical field. Recently, TQ simulations with a realistic multi-injector MGI model have been performed in both {M3D-C1} \citep{kleiner2024extended} and NIMROD \citep{izzo2025disruption} (both without a RE fluid). Both {M3D-C1} and NIMROD indicate that the initial onset of MHD instabilities is dominated by $n=3$ modes due to the injector positions, but $n=1$ modes eventually become dominant \citep{kleiner2024extended,izzo2025disruption}. The {M3D-C1} TQ simulation shows that the field lines become fully stochastic during the TQ and that the period of stochasticity lasts for >\SI{1}{\milli\second}, before flux surfaces reform during the CQ \citep{kleiner2024extended}. NIMROD simulations support this result, also showing fully stochastic field lines during the TQ \citep{izzo2022runaway,izzo2025disruption}. As such, we expect significant losses of any RE seed population during the TQ and RE growth only after closed flux surfaces reform during the CQ, as seen in NIMROD simulations that included RE test particles \citep{izzo2022runaway,izzo2025disruption}.

\section{Summary}

The simulations of SPARC disruptions presented in this paper self-consistently combine runaway electron (RE) generation, 3-D magnetohydrodynamic (MHD) activity, and material injection for the first time. We further self-consistently calculate RE evolution in an axisymmetric "cold" VDE for the first time in a SPARC disruption simulation. The {M3D-C1} simulations demonstrate several effects that arise from RE + MHD coupling, including an initial increase in the RE generation due to the growth of MHD instabilities [see \autoref{fig:unmitigated_3d}\blue{(d-e)}], sawteeth suppression due to increased RE current [see \autoref{fig:unmitigated_3d}\blue{(b,h)}], RE losses due to stochastic field lines (see \autoref{fig:poincare_mitigated}), and RE confinement and plateau formation due to re-healing of flux surfaces (see \autoref{fig:poincare_mitigated} and \autoref{fig:case_C}). As expected from prior experiments and simulations, large RE plateaus are obtained with Ne-only injection, while combined $\text{D}_2$ + Ne injection reduces the RE current (see \autoref{fig:case_D_VDE}). Here, a "cold" VDE is triggered, which terminates the RE beam, preventing formation of a steady plateau, as illustrated in \autoref{fig:case_D_VDE}. A comparison of the VDE with and without REs demonstrates slower vertical motion and later plasma termination with the REs (\autoref{fig:case_D_VDE}\blue{a-c}). These calculations underscore the importance for self-consistently modeling RE and MHD physics during tokamak disruptions, and represent a crucial step in understanding RE generation and mitigation on compact, high-current devices such as SPARC. Future work will extend the present simulations to include higher-fidelity, multi-injector MGI modeling, 3-D non-axisymmetric VDEs, and further investigate how the SPARC REMC can seed and sustain magnetic stochasticity to de-confine REs during the disruption.

\section{Data Availability}

The data that support the findings of this study are available upon reasonable request from the authors.

\section{Declaration of Interests}

The authors have no conflicts of interest to disclose.

\section{Acknowledgments}

This work is supported by Commonwealth Fusion Systems (RPP021: Disruptions). The simulations presented in this paper were performed on the the Engaging cluster at the MGHPCC facility (www.mghpcc.org). This research also used resources of the National Energy Research Scientific Computing Center (NERSC) [10.13039/100017223], a Department of Energy Office of Science User Facility (m3195 and m4829 for year 2024-25). The authors acknowledge J. Chen (PPPL), C. Liu (Peking University), and E. S. Seol (Rensselaer Polytechnic Institute) for their continued support. 






\setcitestyle{square}
\bibliography{main}
\setcitestyle{square}

\end{document}